\documentclass[conference]{IEEEtran}
\usepackage[utf8]{inputenc}
\usepackage[T1]{fontenc}
\IEEEoverridecommandlockouts
\usepackage{cite}
\usepackage{amsmath,amssymb,amsfonts}
\usepackage{graphicx}
\usepackage{textcomp}
\usepackage{xcolor}
\usepackage{multirow}
\usepackage{url}
\def\BibTeX{{\rm B\kern-.05em{\sc i\kern-.025em b}\kern-.08em
    T\kern-.1667em\lower.7ex\hbox{E}\kern-.125emX}}
\begin{document}

\title{A Forensic Analysis of Metadata and Information Leakage in Swedish Public Documents}

\author{
  \IEEEauthorblockN{Noor Abo Hasan \quad Kajsa Linderoth \quad Monia L\"onn}
  \IEEEauthorblockA{\textit{Department of Computer and Systems Sciences, Stockholm University} \\
  Stockholm, Sweden \\
  \textit{All authors contributed equally to this work.}}
}

\maketitle

\begin{abstract}
This study investigates metadata and structural content leakage in publicly available PDF documents from 30 Swedish public-sector authorities. Using an automated collection pipeline and ExifTool-based extraction, 2,876 documents were analyzed and classified using an R0--R3 risk taxonomy. Results reveal that 48.4\% of documents exhibit significant metadata exposure (R2 or R3), with significant variation across institutional tiers ($p < 0.001$). Longitudinal analysis across three legislative periods reveals a statistically significant reduction in metadata exposure between the baseline and post-enactment periods, yet 44.9\% of documents published after the enactment of Sweden's \textit{Cybers\"{a}kerhetslagen} continue to leak sensitive metadata. Furthermore, 41.6\% of documents contain incremental update layers; forensic analysis recovered verifiable text alterations---including deleted personal data and internal identifiers---from 228 of these cases. These findings highlight persistent regulatory lag and a systemic reliance on default, unsanitized document publication workflows within the Swedish public sector.
\end{abstract}


\section{Introduction}
PDF files are widely used for public communication by Swedish government agencies and institutions. Their popularity comes from their ability to preserve document layout and maintain compatibility across different systems, making them a reliable format for publishing official materials such as reports, policy documents, guidelines, and administrative records \cite{iso32000}. In Sweden, this reliance on digital publication is closely connected to the broader transition toward digital governance and the principle of public access to official records, \textit{Offentlighetsprincipen}, which gives the public and media the right to insight into the activities of public authorities \cite{swe_public_access}. This principle supports democratic transparency and accountability \cite{swe_public_access}. However, because large volumes of official documents are made publicly available, publication workflows also become part of the public sector's security posture. If documents are not reviewed or sanitised before publication, they may expose information that was never intended for public access \cite{nsa2008}.

Beyond the content visible to readers, PDF files often contain embedded metadata describing how the file was created, edited, converted, and processed \cite{nsa2008}. This metadata may include author names, usernames, software names and versions, producer fields, and traces of internal workflows \cite{hasegawa2022}. Although this information is usually not visible to ordinary readers, it can be extracted using commonly available forensic and metadata-analysis tools \cite{feng2018}. For public-sector organisations, this makes metadata hygiene an important but sometimes overlooked part of information security \cite{skrauca2026}. Metadata hygiene refers to the practice of reviewing, minimising, or removing unnecessary embedded information before documents are published \cite{spiegel2017}. In a public-sector publishing context, exposed metadata may indicate that document review and sanitisation routines are not applied consistently before publication. Poor metadata hygiene can therefore be viewed as a sign of gaps in organisational information security practices, rather than as an inherent flaw in the PDF format itself. Similar risks have also been observed in other published document formats, showing that hidden data leakage is a broader document-publishing issue rather than a PDF-specific problem \cite{byers2004}.

From a digital forensics perspective, metadata is valuable because it can provide insight into a document's origin, creation process, and later modifications \cite{castiglione2010}. At the same time, when this information is exposed unintentionally, it may create security risks by revealing details about internal systems, user environments, naming conventions, or organisational structures \cite{adhatarao2021}. These details may seem minor on their own, but when collected across many documents, they can support organisational fingerprinting by helping attackers build a broader picture of an organisation's technical environment, software dependencies, document-management practices, and working routines \cite{skrauca2026}. In this way, exposed metadata can expand the attack surface during reconnaissance. For example, staff names or e-mail addresses may enable spear-phishing attempts, while software and version information may help attackers identify specific tools or possible weaknesses to target \cite{adhatarao2021}. Metadata leakage can therefore support more targeted attacks, even when the exposed information appears harmless on its own. Previous research has shown that metadata leakage can support infrastructure fingerprinting and targeted cyber attacks by providing contextual information about both individuals and systems \cite{hasegawa2022}.

The risks are not limited to ordinary metadata fields. A central feature of the PDF format is its support for incremental updates \cite{castiglione2010}. When a PDF is modified, the file is not always rewritten from scratch. Instead, new or changed objects may be appended to the existing file, together with updated cross-reference information and trailer data \cite{castiglione2010}. This design can preserve earlier document states, meaning that deleted, overwritten, or finalised content may remain recoverable unless the document has been properly sanitised, flattened, or regenerated \cite{nsa2008, castiglione2010}. As a result, a publicly available PDF may contain more than its visible final version. It may also include structural traces of earlier edits, hidden content, or previous versions of objects that were assumed to have been removed \cite{garfinkel2014}.

Despite growing awareness of metadata-related risks, limited empirical research has focused specifically on metadata exposure in Swedish public-sector documents. This gap is especially relevant in a highly digitalised environment such as Sweden, where public organisations increasingly rely on digital communication and where regulatory frameworks such as the NIS2 Directive place greater emphasis on cybersecurity, risk management, and organisational accountability \cite{eu-nis2}. Understanding how metadata is exposed in publicly available documents is therefore important not only from a forensic perspective, but also from the perspective of organisational security hygiene and regulatory readiness \cite{skrauca2026}.

\subsection{Research Problem}
The research problem is that publicly available PDFs may contain hidden metadata and structural traces that expose sensitive organisational information, but little is known about the scale, severity, and patterns of this exposure in the Swedish public sector \cite{nsa2008, castiglione2010}. This study therefore investigates whether publicly available PDFs from Swedish public-sector organisations leak metadata or structural traces that may reveal information about internal systems, software environments, user patterns, document workflows, and organisational structures.

This exposure is significant because such information may support organisational fingerprinting when collected across many documents. By combining exposed metadata from multiple files, attackers may build a broader picture of an organisation's technical environment and working routines. This can support reconnaissance, spear-phishing attempts, and other targeted attacks \cite{hasegawa2022, adhatarao2021}.

In addition, metadata exposure reflects broader questions of organisational security hygiene. Public-sector organisations often publish large numbers of documents through different systems and workflows, which increases the risk that embedded information is preserved unintentionally \cite{nsa2008, castiglione2010}. Understanding how often such exposure occurs, what types of information are leaked, and how severe the exposure is can therefore provide insight into both cybersecurity risks and digital forensic implications.

\subsection{Research Question}
This study is guided by the following main research question:

\textit{To what extent, and with what severity, do publicly available PDF documents from Swedish public-sector organisations expose sensitive information through metadata and structural content leakage?}

The following sub-questions support the main research question:
\begin{itemize}
    \item \textbf{RQ1}: How does the prevalence and severity of metadata exposure vary across institutional tiers?
    \item \textbf{RQ2}: How do metadata exposure patterns change across the pre-NIS2, NIS2 preparation, and post-NIS2 implementation periods?
    \item \textbf{RQ3}: To what extent is unintended information recoverable from the incremental update history of publicly available Swedish public-sector PDFs?
\end{itemize}

\subsection{Scope}
This study addresses this gap by analysing metadata and hidden version history in publicly available PDF documents from Swedish public-sector organisations. The analysis examines 2,876 PDF documents collected from 30 Swedish government agencies and institutions in order to identify patterns of information exposure, assess what metadata reveals about internal systems and workflows, and evaluate the implications from both cybersecurity and digital forensic perspectives.

\section{Related Work}
The risks associated with metadata embedded in publicly distributed documents have been documented across several empirical studies. This body of work establishes both the technical mechanisms through which exposure occurs and the organisational conditions under which it persists.

Adhatarao and Lauradoux conducted one of the most technically comprehensive studies in this area, analysing PDF files published by 75 security agencies across 47 countries \cite{adhatarao2021}. Examining 39,664 documents, they found that author information, software environments, and system-related traces were frequently embedded in published files. Critically, their study showed that removing hidden data from PDFs is not straightforward, and that many agencies lacked effective sanitisation routines even where security awareness was otherwise high. This work provides the closest methodological precedent for the present study and establishes the core finding that metadata exposure is widespread even among organisations with explicit security mandates.

Hasegawa, Saito, and Sasaki extended this line of inquiry to Japanese police agencies, analysing 110,989 documents from law enforcement bodies where information security requirements are formally strict \cite{hasegawa2022}. Their results confirmed that metadata leakage occurs even in high-security operational contexts, suggesting that the problem reflects systemic gaps in document workflow management rather than a lack of general security awareness. Their scale and institutional focus make their findings a direct comparison point for the tier-based analysis in the present study.

Within the European public-sector context, Zwaan examined metadata exposure in documents published by Dutch government organisations, finding traces of document creation, editing history, and internal workflows in publicly available files \cite{mulder2017}. Although Sweden differs from the Dutch context---particularly because of Offentlighetsprincipen, which creates a strong legal expectation of public access to official records---Zwaan's work demonstrates that European digital governance environments are not exempt from these risks.

Mendelman examined organisational fingerprinting through public document metadata at Tallinn University of Technology, finding that metadata could reveal internal users, software environments, and operational practices in Estonian governmental organisations \cite{mendelman2018}. More directly comparable to the present study, Skrau\v{c}a analysed approximately 1,000 documents from Estonian governmental, corporate, and educational sectors using the same R0--R3 risk taxonomy applied here \cite{skrauca2026}. Their findings provide a baseline against which the Swedish results can be directly compared, and their classification framework---grounded in NIST SP 800-60---is the methodological foundation adopted in Section~\ref{sec:framework}.

Together, these studies show that metadata leakage in publicly available documents is a recognised security and forensic issue affecting diverse institutional types across multiple countries. However, no prior study has examined Swedish public-sector documents, applied incremental update analysis as a forensic vector, or framed findings against a specific national cybersecurity law. This study addresses all three gaps.

\section{Methodology}
This study employs a mixed-methods design, combining quantitative measurement with qualitative forensic analysis. The quantitative component establishes the prevalence and severity of metadata leakage at scale, enabling statistical comparison across institutional tiers and temporal strata. The qualitative component examines the specific content recoverable from incremental update layers, providing interpretive depth that prevalence rates alone cannot capture. 

\subsection{Methodological Rationale and Alternative Approaches}
The selection of an automated, quantitative forensic analysis was driven by the empirical nature of the research question, which seeks to measure the actual, verifiable extent of information exposure. An alternative methodological approach would have been qualitative, utilizing surveys or semi-structured interviews with IT administrators and publication managers across the targeted authorities to assess their document sanitisation workflows and security awareness. 

While a qualitative approach could provide valuable insight into the \textit{intentions} and internal policies governing document publication, it was rejected for three primary reasons. First, it is highly susceptible to social desirability bias; authorities are likely to over-report compliance with security policies, particularly in a post-NIS2 regulatory environment. Second, qualitative self-reporting measures perceived security posture rather than actual empirical exposure. Third, manual inspection or interviewing cannot scale to provide a statistically significant sample size across 30 distinct institutions. 

Therefore, a quantitative forensic approach was selected as it provides objective, ground-truth evidence of metadata leakage. By directly analysing the published artefacts, this method bypasses self-reporting bias and allows for a reliable, reproducible measurement of actual information exposure at scale.

\subsection{Data Collection}

\subsubsection{Sampling Strategy}
This study employs a two-stage purposive stratified sampling design, operating at both the institutional and document level.

\paragraph{Institutional Sampling (Stage 1)}
At the first stage, 30 Swedish public-sector authorities and institutions were selected through purposive sampling. Rather than random selection, entities were chosen to ensure representation across institutional function, security mandate, and organizational scale. The selected entities were assigned to one of three tiers according to the operational sensitivity of their mandate:

\begin{itemize}
    \item \textbf{Tier 1: Strategic \& Sovereign Risk:} Entities whose primary function concerns national security, intelligence, law enforcement, or executive governance. 
    \item \textbf{Tier 2: Systemic \& Structural Risk:} Entities responsible for critical national infrastructure, financial oversight, and digital governance. 
    \item \textbf{Tier 3: Societal \& Privacy Risk:} Entities that process large volumes of sensitive personal or public-interest data, including healthcare authorities, universities, and social insurance bodies. For a complete list of entities by tier, see Appendix A.
\end{itemize}

\paragraph{Document Sampling (Stage 2)}
At the second stage, documents were collected from each authority using a quota-based sampling frame. 

\paragraph{Temporal Stratification}
The dataset was stratified temporally into three periods aligned with the legislative trajectory of the EU NIS2 Directive \cite{eu-nis2}:
\begin{itemize}
    \item \textbf{Pre-NIS2 (2020--2023):} Baseline period prior to the adoption of the NIS2 Directive.
    \item \textbf{NIS2 Preparation (2024--2025):} Transposition period during which member states were required to implement the directive into national law.
    \item \textbf{Post-NIS2 Implementation (2026--present):} Period following the entry into force of Sweden's national implementation of NIS2, \textit{Cybers\"{a}kerhetslagen} \cite{sfs2025_1506}.
\end{itemize}

\paragraph{Resulting Sample}
The combined sampling procedure yielded 2,876 unique documents acquired across 30 authorities. Table \ref{tab:sample_distribution} summarises the distribution of documents by tier and temporal stratum, excluding 60 documents that fell outside the temporal bounds or had unresolvable dates. These 60 documents were excluded from the temporal comparison but included in the overall results. 

\begin{table}[htbp]
\centering
\caption{Distribution of Unique Documents by Tier and Temporal Stratum}
\label{tab:sample_distribution}
\begin{tabular}{|l|c|c|c|c|}
\hline
\textbf{Tier} & \textbf{2020--2023} & \textbf{2024--2025} & \textbf{2026--Present} & \textbf{Total} \\
\hline
Tier 1 & 378 & 381 & 260 & 1019 \\
Tier 2 & 375 & 393 & 187 & 955 \\
Tier 3 & 298 & 338 & 206 & 842 \\
\hline
\textbf{Total} & \textbf{1051} & \textbf{1112} & \textbf{653} & \textbf{2816} \\
\hline
\end{tabular}
\end{table}

\subsection{Collection Pipeline}\label{sec:pipeline}
A custom automated pipeline was constructed for document discovery, acquisition, and forensic processing:
\begin{itemize}
    \item \textbf{Language:} Python 3.x
    \item \textbf{Libraries:} requests, BeautifulSoup4, hashlib
    \item \textbf{Forensic Tool:} ExifTool (v12.x)
    \item \textbf{Identification:} SHA-256
\end{itemize}

\subsubsection{Automated Discovery}
The primary collection tool was a Python-based crawler using \texttt{BeautifulSoup4} for HTML parsing. The crawler enforced a stay-within-domain policy to ensure data was only collected from targeted servers. Navigation prioritized high-value publication directories, including /publikationer, /rapporter, and /arkiv. For authorities where the main crawler was blocked by WAF infrastructure (Radware or Cloudflare) or where documents were only accessible via JavaScript-rendered search interfaces, dedicated focused scrapers were constructed using direct URL injection or Playwright-based pagination.

\subsubsection{Acquisition and Filtering}
A quota of 50 documents per authority and temporal stratum was enforced to ensure balanced representation across both institutions and time periods, and to prevent high-volume publishers from dominating the dataset. Where an authority did not produce sufficient documents within a given stratum, all available documents were collected without enforcing the quota. Publication year was determined by evaluating, in order of priority: the URL path, the HTTP Last-Modified response header, and the XMP:CreateDate internal metadata field.

\subsubsection{Processing and Integrity}
To maintain a verifiable chain of custody, each acquired file underwent immediate forensic processing upon download. A SHA-256 hash was computed and recorded for every document, serving as a unique identifier and ensuring that all subsequent analysis was performed on the exact byte-sequence as collected. Duplicate files, identified by a matching hash, were excluded from analysis to prevent double-counting.

Metadata extraction was performed using \texttt{ExifTool} (v12.x), a widely adopted command-line tool validated in prior digital forensics literature \cite{adhatarao2021}. \texttt{ExifTool} was invoked per-file in JSON output mode, enabling the extraction of all available metadata fields regardless of schema. The resulting output was normalized and stored in a relational long-format CSV structure, where each row represents a single field--value pair linked to the file's SHA-256 hash, authority, and collection period. This schema enables systematic cross-authority and cross-field comparison.

In parallel, the binary structure of each PDF was analyzed to detect the presence of incremental update layers. Detection was performed by counting occurrences of the \texttt{\%\%EOF} end-of-file marker in the raw byte stream, as each appended revision layer introduces a new cross-reference table and a corresponding \texttt{\%\%EOF} marker. To eliminate false positives, the linearization status of each document was evaluated prior to classification: a linearized PDF inherently contains two \texttt{\%\%EOF} markers as part of its standard structure, and was therefore only flagged if the marker count exceeded two. Non-linearized documents were flagged at a count greater than one. Documents satisfying these criteria were tagged for further forensic examination. 

\subsection{Analytical Framework}\label{sec:framework}

\subsubsection{Exposure Categories}
This study identifies two structurally distinct categories of unintended information exposure in PDF documents. The first is metadata field leakage; structured key-value attributes embedded in the document's Info Dictionary or XMP stream that reveal information about authorship, software, and infrastructure. The second is structural content leakage; the preservation of superseded document versions through the PDF incremental update mechanism. While both categories contribute to the overall attack surface, they differ in nature. While metadata field leakage primarily enables reconnaissance, structural content leakage may expose information the author explicitly intended to remove. These two categories are assessed separately throughout this study.

\subsubsection{Metadata Classification}
In this study, document metadata is treated as a passive information exposure factor. To systematically classify the confidentiality risk posed by each identified metadata field, this study applies the R0--R3 risk taxonomy. The taxonomy is grounded in the risk assessment principles established by NIST SP 800-60 and FIPS 199. NIST SP 800-60 provides a framework for mapping information types to security impact levels based on the potential consequences of unauthorized disclosure, while FIPS 199 defines the three security objectives (confidentiality, integrity, and availability) and establishes Low, Moderate, and High as the standardized impact categories \cite{nist800_60, fips199}. The R0--R3 taxonomy operationalizes these impact levels in a four-point scale adapted specifically for metadata field classification, where each level reflects the degree to which a disclosed field could enable identification, targeting, or infrastructure mapping by an adversary. The following categories were applied:

\begin{itemize}
    \item \textbf{R0 (Non-Sensitive):} Benign structural parameters. File dimensions, PDF version numbers, page counts.
    \item \textbf{R1 (Low Risk):} Broad organizational identities and generic timestamps. Department names, generalized software tools and creation dates without specific timezone mapping.
    \item \textbf{R2 (Medium Risk):} Explicit attribution identifiers mapping to specific internal entities or highly particularized configurations. Specific author names, internal document management IDs, exact patch levels of software, customized printing or pre-press work-flow flags.
    \item \textbf{R3 (High Risk):} Data elements directly charting internal network topographies or system schemas. Absolute file pathways mapped to local C: drives, mapped network shares, internal server IP schemas, organizational Active Directory usernames, or hidden architectural tokens.
\end{itemize}

Classification was automated using a \texttt{Python} script that applied deterministic pattern-matching rules to each extracted field--value pair. Pure filesystem attributes generated by ExifTool during extraction---such as \texttt{FileSize}, \texttt{FilePermissions}, and \texttt{FileModifyDate}---were excluded from classification as they reflect the collection environment rather than embedded document metadata, yielding 66,989 classified fields from the full extraction output. Each document was assigned the highest risk level present among all its classified fields, such that a single R3 field elevated the document's classification to R3 regardless of all other fields present.

\subsubsection{Structural Content Leakage} 
To evaluate the extent to which sensitive information can be recovered from document histories, we utilized a qualitative forensic audit focused on ``Revision Bloat'' and substantive content restoration. Recovery success was measured through a three-stage filter: technical presence of incremental saves, forensic accessibility, and verifiably recovered text alterations (unique ADDED/REMOVED string pairs). To analyze the depth of the exposure, we identified ``outlier'' documents exhibiting extreme revision bloat, defined as documents harboring more than 50 hidden versions.

\subsection{Statistical Framework}

\subsubsection{Tier Comparison}
To quantitatively evaluate the disparity in metadata hygiene across the Swedish administrative landscape, we employed a two-step procedure based on Pearson's Chi-Square ($\chi^2$) tests of independence. First, an omnibus $3 \times 2$ contingency table was constructed, mapping all three organizational Tiers (Independent Variable) against the binary risk classification (Dependent Variable) to test for global variance. Second, a series of post-hoc $2 \times 2$ pairwise comparisons (Tier 1 vs Tier 2, Tier 1 vs Tier 3, and Tier 2 vs Tier 3) were executed to identify specific inter-tier differences.

For the purpose of statistical validity, we established a ``Significant Risk'' threshold by aggregating R2 and R3 occurrences. This aggregation ensured that all cells within the contingency matrix maintained an expected frequency of at least 5, satisfying the fundamental assumptions of the Chi-Square distribution.

The null hypothesis posited that metadata leakage rates are independent of an organization's Tier, while the alternative hypothesis asserted a systemic correlation between organizational mission and digital hygiene. All calculations were performed using Python's \texttt{scipy.stats} library. To correct for the multiple comparison problem inherent in executing three pairwise tests on the same dataset, significance thresholds for the post-hoc tests were adjusted using a Bonferroni correction, setting the required alpha level at $p < 0.0167$.

\subsubsection{Temporal Comparison}
To evaluate the chronological evolution of metadata hygiene relative to the enactment of the Cybers\"{a}kerhetslagen, we employed a two-step procedure consistent with our sectoral analysis. First, an omnibus $3 \times 2$ Chi-Square test of independence was executed, mapping the three distinct temporal epochs (Pre-NIS2, NIS2 Preparation, and Post-Cybers\"{a}kerhetslagen) against the binary R2+ risk classification. Second, post-hoc $2 \times 2$ pairwise comparisons were conducted between the temporal epochs to isolate specific periods of statistical difference.

Consistent with our sectoral analysis, we utilized the aggregated ``Significant Risk'' (R2+) threshold to satisfy the Chi-Square requirement for expected cell frequencies ($E \geq 5$). The null hypothesis posited that metadata exposure rates are independent of the document publication period. To control for the multiple comparison problem during post-hoc testing, significance thresholds were adjusted using a Bonferroni correction, setting the required alpha level at $p < 0.0167$.

\subsection{Ethical Considerations}
This study examines only PDF documents that were publicly available through official websites and online archives at the time of collection. Public availability does not remove the responsibility to handle data carefully, particularly where metadata may contain names, usernames, software details, or other organisational information.

All collected PDF files and extracted metadata were stored locally on encrypted workstations and will be deleted upon completion of the research. Personal names, usernames, and specific technical identifiers are not reproduced in the reported findings, and individual authorities are referred to in the results only by tier-based labels (see Section~\ref{sec:tier}). The collected files and any sensitive extracted metadata will not be redistributed.

Data collection was conducted to minimise impact on targeted servers. The collection script observed each domain's robots.txt file, restricted crawling to permitted paths, and applied a random request delay of 3.0--5.0 seconds. The analysis scripts are available on GitHub for transparency and replication. Any processing of personal data identified during the study is conducted under the research exemption provided by GDPR Article 89, which permits the processing of personal data for scientific research purposes subject to appropriate safeguards \cite{eu-gdpr}. The affected organizations will be notified, and specific identifiers have been removed. 

\section{Results}

\subsection{Overall Risk Distribution}
In accordance with the metadata risk taxonomy defined in the Methodology, all 66,989 extracted metadata fields were classified from R0 to R3. At the document level, 43.0\% of documents exhibited R2 (Medium Risk) as their highest risk factor, and 5.4\% exhibited R3 (High Risk). 

\begin{figure}[htbp]
    \centering
    \includegraphics[width=\linewidth]{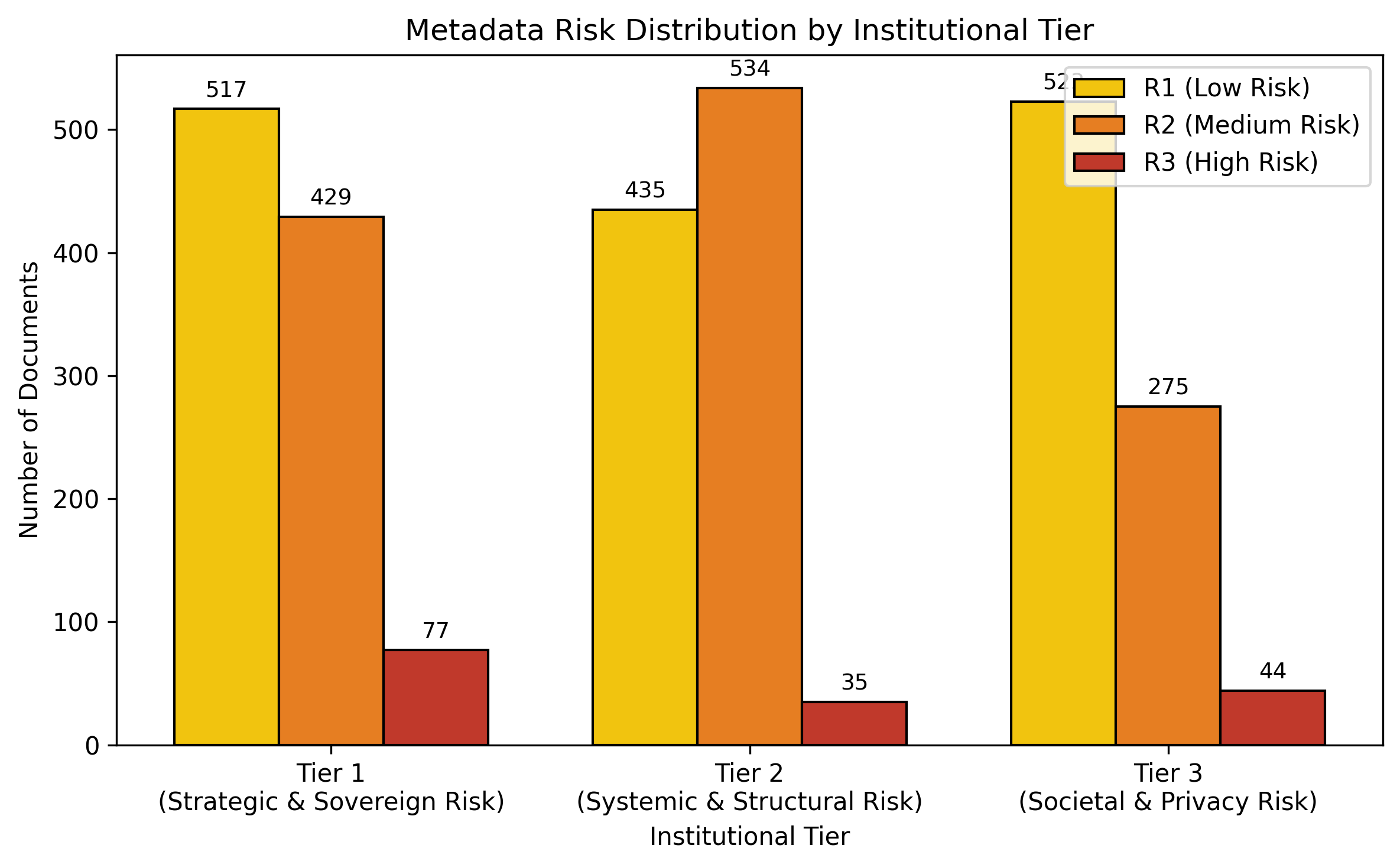}
    \caption{Metadata Risk Distribution by Institutional Tier.}
    \label{fig:tier_risk}
\end{figure}

Figure \ref{fig:tier_risk} presents the distribution of R1, R2, and R3 risk levels across the three institutional tiers.

\subsection{Types of Information Exposed}
Beyond broad risk classifications, it is important to understand exactly what types of sensitive information are being exposed in publicly available public-sector documents. To analyze this, the specific leaked metadata fields were grouped into five categories. Since a single document can leak multiple types of information (e.g., both a personal name and an email address), the percentages reflect the proportion of the total unique documents that contained at least one instance of that specific data type.

\begin{figure}[htbp]
    \centering
    \includegraphics[width=\linewidth]{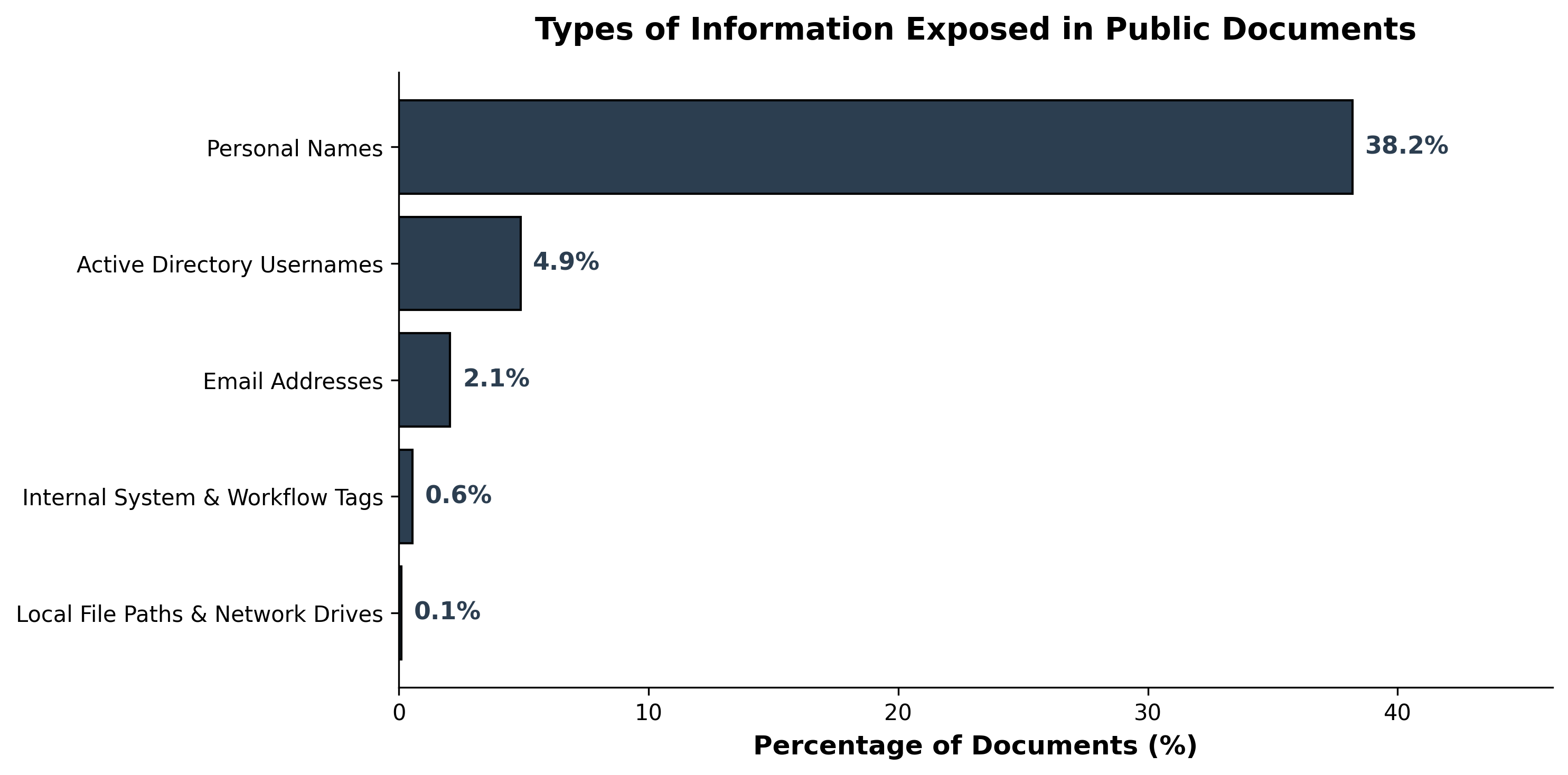}
    \caption{Percentage of documents exposing specific types of sensitive information. Categories are non-exclusive.}
    \label{fig:data_type_exposure}
\end{figure}

Figure \ref{fig:data_type_exposure} visualizes this breakdown. Personal names were the most frequently exposed sensitive data type, found in 38.2\% of all analyzed documents. Active Directory Usernames, which map directly to internal authentication schemas, were exposed in 4.9\% of documents. Email addresses were found in 2.1\% of documents. More critically, 0.6\% of documents leaked internal system and workflow tags (such as Microsoft Information Protection labels that expose internal Tenant IDs), and 0.1\% leaked absolute local file paths or mapped network drives.

\subsection{Sectoral Tier Comparison}\label{sec:tier}

The omnibus Chi-Square test revealed a highly significant global difference in R2+ exposure across the three tiers ($\chi^2(2) = 63.75, p < 0.001$). As detailed in Table \ref{tab:chi_tier}, post-hoc pairwise comparisons confirm that Tier 2 (Systemic \& Structural Risk) exhibited the highest rate of significant risk exposure (56.7\%), followed by Tier 1 (Strategic \& Sovereign Risk, 49.5\%) and Tier 3 (Societal \& Privacy Risk, 37.9\%). All three pairwise comparisons yielded statistically significant results that comfortably survive the strict Bonferroni-corrected threshold ($p < 0.0167$), allowing us to reject the null hypothesis across all combinations.

\begin{table}[htbp]
\centering
\caption{Chi-Square Test Results: Sectoral Tier Comparison (R2+ Risk)}
\label{tab:chi_tier}
\begin{tabular}{|l|c|c|c|c|}
\hline
\textbf{Comparison} & \textbf{R2+ Exposure} & \textbf{$\chi^2$} & \textbf{p-value} & \textbf{Sig.} \\
\hline
Tier 1 vs. Tier 2 & 49.5\% vs 56.7\% & 10.29 & 0.0013 & * \\
Tier 1 vs. Tier 3 & 49.5\% vs 37.9\% & 24.62 & < 0.0001 & * \\
Tier 2 vs. Tier 3 & 56.7\% vs 37.9\% & 62.38 & < 0.0001 & * \\
\hline
\multicolumn{5}{l}{\footnotesize Omnibus: $\chi^2(2) = 63.75, p < 0.001$} \\
\multicolumn{5}{l}{\footnotesize * Significant at the Bonferroni-corrected threshold of $\alpha = 0.0167$} \\
\end{tabular}
\end{table}

While the Chi-Square analysis establishes statistical trends at the aggregate tier level, examining individual authorities reveals significant intra-tier variance. To avoid directing attention to specific organisations and documents, individual authorities are referred to using labels that indicate only their tier (e.g., Authority T1-A). Labels are assigned in order of first appearance and are used consistently throughout the remainder of the paper. Table \ref{tab:auth_extremes} highlights the extreme ends of the metadata hygiene spectrum by showcasing the best and worst performing authorities within each tier, measured by their percentage of documents exhibiting significant risk (R2 and R3).

Notably, even within Tier 1 (Strategic \& Sovereign Risk), performance ranges dramatically from Authority T1-A, which exhibited excellent metadata hygiene with only 13.8\% exposure, to Authority T1-B, which leaked significant risk data in 92.0\% of its published documents. Similarly, in Tier 3, Authority T3-A achieved high metadata hygiene with only 6.6\% exposure across its sampled documents.

\begin{table}[htbp]
\centering
\caption{Metadata Hygiene Performance Extremes by Tier (R2+R3 Exposure)}
\label{tab:auth_extremes}
{\small
\begin{tabular}{|l|l|p{3.2cm}|c|}
\hline
\textbf{Tier} & \textbf{Perf.} & \textbf{Authority} & \textbf{Exp.} \\
\hline
\multirow{2}{*}{T1} & Best & Authority T1-A & 13.8\%  \\
 & Worst & Authority T1-B & 92.0\% \\
\hline
\multirow{2}{*}{T2} & Best & Authority T2-A & 14.8\%  \\
 & Worst & Authority T2-B & 90.5\%  \\
\hline
\multirow{2}{*}{T3} & Best & Authority T3-A & 6.6\% \\
 & Worst & Authority T3-B & 83.6\% \\
\hline
\end{tabular}
}
\end{table}

\subsubsection{Notable Forensic Findings}
Authority T1-A's exposure consisted almost entirely of software version identifiers (R2). One exception was a template path referencing another authority's internal server---a third-party trace embedded in a shared template rather than in the authority's own publishing pipeline.

Authority T1-B exhibited the highest personal name exposure of any authority in the dataset, with 85 unique individuals identifiable through the Author field alone. Despite its low incremental update rate, one document's binary history contained a removed internal document-handling field and an internal production instruction, both absent from the published version but recoverable through forensic extraction.

Authority T1-C exposed an internal file path in one document, alongside an individual author name. Across its sample, 45 Author field entries contained internal case reference numbers.

Authority T1-D's documents contained a high number of R3-classified EmbeddedHref fields, including internal directory and file references that originate from a predecessor authority, dissolved in an earlier reorganisation, and that persist in current templates. One Author entry combined an individual's name with an internal unit identifier.

Despite Authority T3-A's low overall exposure, a single embedded field in one of its documents exposed an internal infrastructure URL.

Across the dataset, third-party platform identifiers from Mendeley, Grammarly, and Articulate Storyline were found embedded in documents from authorities with otherwise clean metadata profiles.

\subsection{Temporal Comparison}

\begin{figure}[htbp]
    \centering
    \includegraphics[width=\linewidth]{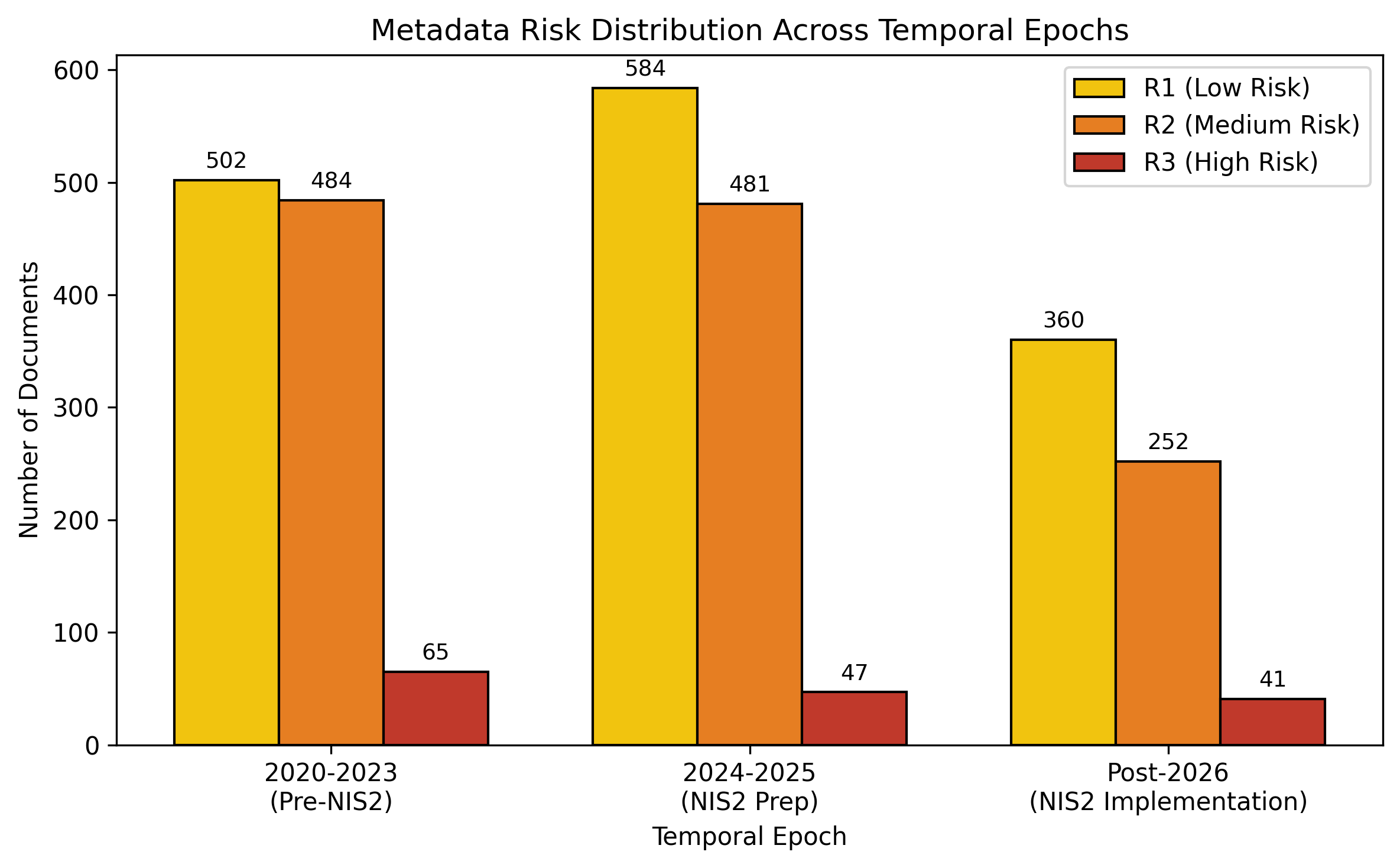}
    \caption{Metadata Risk Distribution Across Temporal Epochs.}
    \label{fig:temporal_risk}
\end{figure}

Figure \ref{fig:temporal_risk} visualizes the risk distribution across the three temporal epochs. Table \ref{tab:chi_temporal} presents the results of the $3 \times 2$ Chi-Square test across all three periods.

As shown in Table \ref{tab:chi_temporal}, the omnibus $3 \times 2$ Chi-Square test revealed a statistically significant difference in R2+ risk exposure across the three temporal epochs ($\chi^2(2) = 9.75, p = 0.0076$). Exposure dropped from 52.2\% during the Pre-NIS2 baseline (2020--2023), to 47.5\% during the NIS2 Preparation period (2024--2025), and 44.9\% after the enactment of the Cybers\"{a}kerhetslagen (Post-2026). 

However, post-hoc pairwise comparisons reveal that this reduction is only statistically significant when comparing the baseline directly to the post-enactment period ($\chi^2 = 8.45, p = 0.0036$). The interim reductions between Pre-NIS2 and NIS2 Prep ($p = 0.0302$), and between NIS2 Prep and Post-CSL ($p = 0.3111$), do not survive the strict Bonferroni-corrected threshold ($p < 0.0167$). 

Furthermore, even the most recent epoch exhibits an exposure rate of 44.9\%, indicating that nearly half of all documents published after the enactment of the Cybers\"{a}kerhetslagen continue to leak significant metadata.

\begin{table}[htbp]
\centering
\caption{Chi-Square Test Results: Temporal Comparison (R2+ Risk)}
\label{tab:chi_temporal}
\begin{tabular}{|l|c|c|c|c|}
\hline
\textbf{Comparison} & \textbf{R2+ Exposure} & \textbf{$\chi^2$} & \textbf{p-value} & \textbf{Sig.} \\
\hline
Pre-NIS2 vs. Prep & 52.2\% vs 47.5\% & 4.70 & 0.0302 & n.s. \\
Prep vs. Post-CSL & 47.5\% vs 44.9\% & 1.03 & 0.3111 & n.s. \\
Pre-NIS2 vs. Post & 52.2\% vs 44.9\% & 8.45 & 0.0036 & * \\
\hline
\multicolumn{5}{l}{\footnotesize Omnibus: $\chi^2(2) = 9.75, p = 0.0076$} \\
\multicolumn{5}{l}{\footnotesize * Significant at the Bonferroni-corrected threshold of $\alpha = 0.0167$} \\
\multicolumn{5}{l}{\footnotesize n.s. = not significant} \\
\end{tabular}
\end{table}

\subsection{Structural Content Leakage}
To address RQ3, the study evaluated the prevalence of incremental update layers across the dataset using the linearization-aware detection method defined in Section~\ref{sec:pipeline}.

\subsubsection{Prevalence}
The analysis revealed that structural content leakage is a significant phenomenon across the Swedish public sector, with 41.6\% (1,196/2,876) of all sampled documents containing at least one appended incremental update layer beyond their standard structural requirements.

\subsubsection{Revision Bloat and Outlier Analysis}
The depth of structural leakage was measured through ``Revision Bloat,'' representing the total number of historical document states preserved within the binary stream. While the median document harbored only 2--3 versions, the study identified extreme outliers where document histories spanned dozens of internal revisions.

Three documents were identified as ``extreme outliers'' harboring more than 50 versions:
\begin{itemize}
    \item \textbf{Authority T3-C (annual report)}: 130 versions.
    \item \textbf{Authority T3-C (annual report, preceding year)}: 123 versions.
    \item \textbf{Authority T3-D (statistical report)}: 84 versions.
\end{itemize}

\subsubsection{Content Restoration and Verifiable Text Alterations}
While the technical presence of incremental updates was significant (1,196 documents), the forensic accessibility of previous content varies. To assess the severity of this leakage, a deep forensic extraction was performed. The analysis successfully recovered verifiable text alterations (unique ADDED/REMOVED string pairs) from 228 documents. This demonstrates that for nearly 8\% of the entire dataset, substantive edits, deleted text, or prior drafts remain accessible to the public.

Three representative cases illustrate the range of recoverable content:

\begin{itemize}
    \item \textbf{Personal Data Handling (Authority T1-E):} In an archive description document, Version 2 removed descriptions of who handles personal data processing, including references to internal IT administrator roles and local office units. This operational detail about data processing responsibilities remains recoverable.
    \item \textbf{Individual Attribution (Authority T1-F):} In a regulatory publication, the named publisher and associated ISSN metadata were removed before the final version. The publisher's name remains recoverable from the document's incremental history.
    \item \textbf{Legal Text Restructuring (Authority T3-E):} A government form underwent 36 revisions. Version 34 contained a full prose paragraph listing all legal entity types covered by a provision, which was later restructured into a condensed bullet list. The original, more expansive legal language is fully recoverable.
\end{itemize}

Authority T1-G was a notable exception, with only 3 of its 121 sampled documents (2.5\%) containing incremental saves beyond linearization requirements.

\section{Analysis and Discussion}
The results demonstrate that metadata leakage in Swedish public-sector PDF documents is both pervasive and measurable, with patterns that are systemic rather than incidental.

\subsection{RQ1 --- Institutional Tier and Metadata Hygiene}
The chi-square analysis confirms a statistically significant difference in R2+ exposure across tiers, suggesting that institutional mandate is associated with metadata hygiene. However, the aggregate tier result conceals substantial intra-tier variance that the classification alone cannot explain.

Authority T1-A's near-exclusive exposure of software version identifiers demonstrates what effective pipeline management can achieve, yet the presence of a single third-party template path shows that even well-managed pipelines can carry residual traces from cross-agency collaboration.

Authority T1-B most directly challenges the tier model. Despite its Tier 1 mandate, it exhibits the highest personal name exposure of any institution---85 unique individuals through the Author field alone. Its structural profile adds a further dimension: despite this extreme metadata exposure, a comparatively lower share of its documents contained actionable incremental saves, suggesting the use of PDF converters that flatten structural history but fail to strip embedded metadata tags. This decoupling of the two leakage vectors underscores that sanitisation must address both metadata fields and document structure independently.

Authority T1-C's exposed internal file path and the 45 internal case reference numbers in its Author fields illustrate how administrative metadata conventions can inadvertently link published documents to internal governance processes.

The legacy artefacts at Authority T1-D, inherited from a dissolved predecessor authority, demonstrate that template inheritance across organisational transitions is a distinct and persistent vector that standard sanitisation routines do not address.

Institutional mandate is therefore associated with differences in aggregate exposure, but a strong security mandate does not in itself ensure good metadata hygiene, and tier is an insufficient predictor at the document level. Authority T3-A illustrates the same point from the opposite direction: despite its low overall exposure, a single internal infrastructure URL may enable reconnaissance even within an otherwise sanitised portfolio.

Beyond standard metadata fields, identifiers from Mendeley, Grammarly, and Articulate Storyline enable cross-document attribution across platforms---a class of exposure that standard sanitisation tools, typically configured for well-known fields, are unlikely to detect.

\subsection{RQ2 --- Temporal Patterns and the NIS2 Trajectory}
The temporal analysis reveals a statistically significant reduction in R2+ exposure between the Pre-NIS2 baseline and the Post-NIS2 enactment period. However, the Post-Cybers\"{a}kerhetslagen stratum contains substantially fewer documents than the preceding periods, as fewer publications had been released within this shorter window at the time of collection. The observed reduction may therefore reflect this compositional difference in addition to---or instead of---a genuine behavioural response to regulatory pressure. Future replication with a larger Post-Cybers\"{a}kerhetslagen sample is necessary to confirm whether the reduction is durable.
Even interpreted conservatively, the 44.9\% post-enactment exposure rate indicates that legislative compliance does not translate automatically into document-level sanitisation practices. While \textit{Cybers\"{a}kerhetslagen} enforces systemic risk management and incident reporting, it prescribes no specific operational standards for document publication workflows. Consequently, the regulatory pressure creates a statistically visible improvement at the macro level, but remains practically insufficient to close granular leakage vectors.

\subsection{RQ3 --- Recoverability of Unintended Information from Incremental Updates}
Of the 1,196 documents containing true incremental update layers, forensic extraction identified verifiable content changes in 228. The majority were editorial, such as correcting spelling mistakes, whitespaces or bullet points, carrying limited forensic significance in isolation. However, among the 228, 27 contained substantive removals---spanning personal data handling descriptions, named individuals, and internal production instructions---demonstrating that the incremental update mechanism can preserve precisely the information that authors intended to retract.

The 41.6\% prevalence rate suggests that ``saving'' rather than ``exporting'' is a common publication workflow, and that many organisations do not flatten or regenerate PDFs before publication. The extreme outliers illustrate the depth of this problem: a document containing 130 version markers means that each point where a user modified the file---correcting financial figures, changing phrasing, or removing sensitive placeholders---remains preserved in the binary stream because no final flattening or sanitisation was applied.

These findings must be contextualised within the scope of the study. The documents examined were published for public access and do not in most cases contain highly sensitive visible content. However, the same mechanism applies to documents released on request under Offentlighetsprincipen, where redaction targets the visible layer while the incremental history may remain intact. The 41.6\% prevalence of unsanitized PDFs indicates that many authorities do not apply flattening before release---a direct implication for cases where redaction is not merely cosmetic, and an important direction for future research. When an organisation removes content but that information remains recoverable in the binary stream, it highlights a critical gap: Swedish public authorities are largely relying on the PDF format's convenience without understanding its binary persistence.

Authority T1-G provides a positive counterexample: only 3 of its 121 sampled documents (2.5\%) contained incremental saves beyond linearization requirements, demonstrating that centralised, sanitised PDF export workflows that strip revision history are operationally achievable.

\section{Limitations and Future Work}
This study is limited to documents publicly available through official websites and online archives at the time of collection, explicitly excluding unpublished files and internal systems. The scope focuses strictly on PDF metadata and structural content leakage, rather than visible textual content or an organization's broader cybersecurity posture. Methodologically, the purposive sampling frame implies that documents originating from the same authority share publication workflows, leading to intra-authority clustering rather than perfect statistical independence---a standard limitation in large-scale document analysis. The R0--R3 classification was applied programmatically across 66,989 metadata fields; while the taxonomy is grounded in established risk principles, automated classification at this scale may introduce false positives and false negatives, and the timeframe of this study did not permit manual review and reclassification of all fields. Additionally, the Post-NIS2 temporal stratum is smaller and covers a shorter period than preceding strata; therefore, the observed downward trend in exposure must be replicated with a larger sample before a durable shift can be confirmed.

Future research should investigate whether the unsanitized publication pipelines identified here also apply to documents released upon request under Offentlighetsprincipen, where visible redactions may inadvertently leave the incremental update history intact and recoverable. Furthermore, future work should assess the effectiveness of automated metadata sanitization tools within the Swedish public sector and expand the research scope to municipal-level authorities to uncover further patterns of exposure.
\section{Conclusion}
This study demonstrates that metadata leakage in publicly available Swedish public-sector PDFs is both pervasive and systemic. Across 2,876 documents from 30 authorities, sensitive information---including personal names, Active Directory usernames, internal file paths, and third-party platform identifiers---was recoverable through standard forensic extraction. Institutional tier is associated with hygiene differences but is an insufficient predictor at the document level; the evidence points to publication workflow design as the primary driver of exposure. The structural content analysis confirms that content explicitly removed before publication remains recoverable from the binary stream. The positive outliers in the dataset confirm that thorough sanitization is operationally achievable, making this a policy and workflow problem rather than a technological one.
\section*{Code and Data Availability}
To support reproducible research, the custom automated forensic pipeline developed for this study---including the collection scrapers, metadata extraction logic, structural analysis scripts, and statistical test implementations---has been open-sourced. The source code is available at \url{https://github.com/Nooorab/PDFMETA-Replication-Package}. Due to the potentially sensitive nature of the extracted information (e.g., internal infrastructure paths and Active Directory usernames), the raw PDF dataset and the resulting metadata database have been withheld to prevent unnecessary exposure.

\bibliographystyle{IEEEtran}
\bibliography{references}

\newpage
\section*{Appendix A: Institutional Sampling Frame}
The following 30 Swedish public-sector authorities and institutions were selected for analysis, categorized by their primary operational mandate into three tiers of risk.

\subsection*{Tier 1: Strategic \& Sovereign Risk}
\begin{itemize}
    \item Regeringskansliet
    \item Försvarets radioanstalt (FRA)
    \item Försvarsmakten
    \item Myndigheten för samhällsskydd och beredskap (MSB)
    \item Myndigheten för säkerhets- och integritetsskydd
    \item Tullverket
    \item Totalförsvarets forskningsinstitut (FOI)
    \item Polismyndigheten
    \item Åklagarmyndigheten
    \item Säkerhetspolisen (SÄPO)
\end{itemize}

\subsection*{Tier 2: Systemic \& Structural Risk}
\begin{itemize}
    \item Myndigheten för digital förvaltning (DIGG)
    \item E-hälsomyndigheten
    \item Integritetsskyddsmyndigheten (IMY)
    \item Ekobrottsmyndigheten
    \item Energimyndigheten
    \item Lantmäteriet
    \item Affärsverket svenska kraftnät
    \item Post- och telestyrelsen (PTS)
    \item Trafikverket
    \item Transportstyrelsen
\end{itemize}

\subsection*{Tier 3: Societal \& Privacy Risk}
\begin{itemize}
    \item Karolinska institutet
    \item Lunds universitet
    \item Socialstyrelsen
    \item Folkhälsomyndigheten
    \item Statistiska centralbyrån (SCB)
    \item Stockholms universitet
    \item Bolagsverket
    \item Pensionsmyndigheten
    \item Försäkringskassan
    \item Skatteverket
\end{itemize}

\end{document}